\documentclass{article}
\usepackage{graphics}
\begin{document}
\font\smallerfont cmr8
\def\enew{E_{\rm new}}
\def\ler{\langle E\rangle}
\def\muoneN{\mu_{1,N}}
\def\mutwoN{\mu_{2,N}}
\def\nri{N\rightarrow\infty}
\def\eold{E_{\rm old}}
\def\xu{\underline x}
\def\xiu{\underline \xi}
\hbadness=16000
\vbadness=16000
\overfullrule=0pt

\title{Conditional Expectations and Renormalization}

\author{
Alexandre J.\ Chorin \\
Department of Mathematics\\
University of California\\
Berkeley, CA 94720
}
\date{}

\maketitle

\begin{abstract}
In
optimal prediction methods one estimates the future behavior of 
underresolved systems
by solving reduced systems of equations for expectations conditioned by partial data; 
renormalization group methods reduce the number of variables in complex systems 
through integration of unwanted scales.
We establish the relation between these methods for systems in 
thermal equilibrium,
and use this relation to find renormalization parameter
flows
and the coefficients in reduced systems by expanding conditional expectations in series and evaluating the coefficients 
by Monte-Carlo.
We illustrate the construction by finding parameter flows  
for  simple spin systems and then using the renormalized (=reduced) systems to calculate the critical temperature
and the magnetization.
\end{abstract}


\vfill\eject
\baselineskip14pt
\noindent
\section{Introduction}

In the optimal prediction (OP)
methods presented in earlier work by the author and others
\cite{CKK1},\cite{CHK},\cite{CKL},\cite{Chorin7},
an estimate of the
future solution of an underresolved problem,
or of a problem where the initial data are only partially known, was obtained
by solving a reduced system of equations for the conditional expectation of
the solution given the
partial data.
This system,  
closely related to a generalized Langevin equation of the Mori-Zwanzig type \cite{evans},\cite{zwanzig2},
is derived in detail in \cite{CHK3}.
Short-time estimates can be obtained by keeping only the first term
on the right hand side of this system, obtaining a relation between
the rate of change of a reduced set of
variables and conditional expectations of
the full rate of change; a simplified derivation of this relation is given below. 
Hald's theorem \cite{CHK} asserts 
that if one starts with a Hamiltonian system, then the reduced system 
obtained in this way 
is also Hamiltonian,
with a Hamiltonian equal to a conditional free energy of the
original Hamiltonian system. 

Renormalization group (RNG) transformations \cite{burkhardt},\cite{kadanoff},\cite{ma}
reduce the dimensionality of a system of equations
by integrating out
unwanted scales.
That there is a qualitative resemblance between OP and RNG
methods is quite clear, and
has been pointed out in particular in the related work of Goldenfeld et al. \cite{goldenfeld},\cite{hou}.
In the present paper we focus on the special case of
Hamiltonian systems in thermal equilibrium, and show that in this case the
RNG transformations of the
Hamiltonian can be obtained by integrating conditional expectations of the
derivatives of the
Hamiltonian; loosely speaking, RNG transformations are integrals of OP
reductions.
This remark, based on Hald's theorem,  makes
possible the efficient evaluation of the coefficients of the new Hamiltonians in RNG transformations
(the \lq\lq{RNG parameter flow}")
by simple Monte-Carlo methods, for example
by Swendsen's small-cell Monte-Carlo RNG \cite{burkhardt},\cite{swendsen}.
The coefficients in the new Hamiltonian define the reduced system of equations used to estimate 
the future in OP.
To illustrate the construction, we apply
it to spin systems and obtain explicitly the parameter flows in addition to 
critical points, critical exponents, and order parameters.
We exhibit in detail a particular implementation that is a little awkward
if viewed as an instance of a RNG
but is particularly convenient for OP.

A little thought shows that what is offered in the present
paper is a numerical
short-cut. Suppose $x=(x_1,x_2,\ldots)$ is a set of $n$ random variables
($n$ may be infinite),
and let $m<n$; partition $x$ so that $x=(\hat{x},\tilde{x})$,
$\hat{x}=(x_1,x_2,\ldots,x_m), \tilde{x}=(x_{m+1},x_{m+2},\dots)$.
Let $p=p(x)$ be the joint probability density of all the variables,
and consider the
problem of finding a function $\hat{H}=\hat{H}(\hat{x})$ such that 
$$
\exp(-\hat{H}(\hat{x}))=\int p(x)d\tilde{x},
\eqno(1)
$$
where $d\tilde{x}=dx_{m+1}dx_{m+2}\cdots$.
There is no question that $\hat{H}$ is well defined but the obvious ways of
finding it can  be costly.
We are offering effective ways to do so. There are other situations where one
wants to integrate out unwanted variables inside nonlinear functions,
and our short-cut may serve there as well;
in subsequent papers we shall apply it to problems in
irreversible statistical mechanics and, equivalently, to problems involving
the full long-time OP equations. 

\vskip14pt
\noindent
\section{Conditional expectations and optimal prediction}

Consider a set $x$ of random variables $(x_1,x_2,\ldots,x_n)$ with a joint
probability density of the form $Z^{-1}e^{-H(x)}$,
$Z=\int e^{-H(x)}dx$, $dx=dx_1dx_2\ldots dx_n$.
Consider the space $L_2$ of function $u(x), v(x),\ldots$,
with the inner product $\langle u,v\rangle=E[uv]=\int u(x)v(x)Z^{-1}\exp (-H)dx$, where
$E[\cdot]$ denotes an expected value.

Partition the variables into two groups as above,
$x=(\hat x ,\tilde x)$, $\hat x=(x_1,\ldots,x_m)$, $m<n$.
Given a function $f(x)$, its conditional expectation given $\hat{x}$ is
$$E[f(x)|\hat x]=
{\displaystyle \int f(x) e^{-H(x)} d\tilde x
\over\displaystyle\int e^{-H(x)}d\tilde x}; 
\eqno(2)$$
it is the average of $f$ keeping $\hat x$ fixed.
The conditional expectation is a function of $\hat x$ only,
and it is the best approximation of $f$ in the mean square sense
by a function of $\hat x$:
$$E\left[\left(f(x)-E[f(x)|\hat x]\right)^2\right]
\leq E\left[\left(f(x)-h(\hat x)\right)^2\right]\eqno(3)$$
for any function 
$h=h(\hat x)$.
$E[f|\hat x]$ is the orthogonal projection of $f$ onto the subspace 
$\hat L_2$ of $L$ that contains functions of $\hat x$ only.
$E[f(x)|\hat x]$ can be approximated by expansion in a basis of $\hat L_2$; 
keeping only a suitable finite number $\ell$ of basis functions 
$\varphi_1(\hat x),\varphi_2(\hat x),\ldots,\varphi_\ell(\hat x)$,
and minimizing the distance between $f$
and the span of the $\varphi_i(\hat x)$,
one finds 
$$E[f|\hat x]=\sum_{i=1}^\ell c_i \varphi_i (\hat x),$$
where the $c_i$ satisfy the equation
$$\Phi c=r,\eqno(4)$$
where $\Phi$ is the matrix with elements
$\Phi_{ij}=\langle \varphi_i,\varphi_j\rangle$, $c=(c_1,\ldots,c_\ell)$,
and $r=\left(\langle f,\varphi_1\rangle,\langle f,\varphi_2\rangle\ldots,\langle f,\varphi_\ell\rangle\right)$.
Usually the inner products can be calculated by Metropolis sampling.

Suppose you want to find a function  $\hat H=\hat H(\hat x)$ such that 
$$
e^{-\hat H(\hat x)}=\int e^{-H(\hat x,\tilde x)} d\tilde x,\eqno(5)
$$
i.e., write the marginal probability density of the variables $\hat x$ in
exponential form.
Suppose one can write 
$$H(x)=\sum_{i=1}^\ell \alpha_i \varphi_i (x),$$
and 
and let 
$i\leq m$, where $m$ is the number of components of  the vector $\hat x=$
Then
$$\matrix{
E\left[{\displaystyle\partial\over \displaystyle\partial x_i} H(x)|\hat x\right]
&=&
{\displaystyle\int {\displaystyle\partial\over \displaystyle\partial x_i}\displaystyle H(x)e^{-H(x)} d\tilde x\over\displaystyle\int e^{-H(x)}d\tilde x}\hfill\cr
&&\cr
&=&{\displaystyle\partial\over \displaystyle\partial x_i}\left(-\log\int e^{-H(x)}d\tilde x\right).\hfill}\eqno(6)$$
An analogous relation between the derivative of a logarithm of a partially integrated density and a conditional
expectation arises also in the context of expectation-maximization is statistics \cite{bickel2}. 

If one can find a basis for $\hat L_2$ consisting of functions of the form
${\displaystyle\partial\over\displaystyle\partial x_1}\varphi_j (\hat x)$, $j=1,\ldots,$ and provided 
the set of variables is homogenous so that for all $i\leq m$
the coefficients $c_j$ in the expansions 
$\left[ {\displaystyle\partial\over\displaystyle\partial x_i} H|\hat x\right]
=\sum_{j=1}^\ell c_j{\displaystyle\partial\over\displaystyle\partial x_i}\varphi_j(\hat x)$ 
are independent of $i$, then the expansion
$$\hat H(\hat x)=\sum c_j\varphi_j(\hat x).\eqno(7)$$
follows immediately.
This is our key observation.

This construction is just Hald's theorem for OP \cite{CHK}:     
Suppose one has a system of 
differential equations (written as ordinary differential equations for
simplicity) of the form
$${\displaystyle d\over \displaystyle dt}\varphi(t)=R\left(\varphi(t)\right),
\ \ \varphi(0)=x\eqno (8)$$
where $\varphi, R$ and $x$ are $n$-vectors with components
$\varphi_i,R_i,x_i,$ $i=1,\ldots,n$ and $t$ is the time.
Suppose we partition as above $\varphi=(\hat \varphi,\tilde\varphi)$,
$R=(\hat R,\tilde R)$,
where $\hat \varphi$ contains the first $m$ components of $\varphi$, etc.
Suppose the system (8) is Hamiltonian, i.e.,
$m,n$ are even,
$R_i={\displaystyle\partial\over\displaystyle\partial x_{i-1}} H$
for $i$ even, 
$R_i=-{\displaystyle\partial\over\displaystyle\partial x_{i+1}} H$ for $i$ odd; $H=H(x)$ is
the Hamiltonian and 
$Z^{-1}e^{-H}$ is then an invariant probability density for the system.

Suppose we can afford to solve only $m<n$ of the equations in (8) or have only
$m$ data components $\hat x$. 
We want to solve equations for $\hat \varphi$:
$${\displaystyle d\hat \varphi\over \displaystyle dt}=\hat R(\varphi), \ \ \hat\varphi(0)=\hat x,$$
where $i\leq m$,   
but the argument of $\hat R$ is the whole vector $\varphi$.
It is natural to approximate $\hat R_i(\varphi)$ by the closest function 
of 
$\hat \varphi$ for each $i\leq m$, i.e., solve
$${\displaystyle d\hat\varphi\over\displaystyle dt}=E[\hat R(\varphi)\mid \hat\varphi].\eqno(9)$$

The approximation (9) is valid only for a short time, as one can see from the
full equation for the evolution of  
${\displaystyle d\hat\varphi\over\displaystyle dt}$ in \cite{CHK},\cite{CHK3}, see 
also below.
Hald's theorem asserts that the system (9) is Hamiltonian, with Hamiltonian
$\hat H=\hat H(\hat x)=-\log\int e^{-H(x)}d\tilde x$, a relation identical to 
equation (6). 
The existence of $\hat H$ shows that the approximation (9)  cannot be valid
for long times:
the predictive power of partial initial data decays at $t\rightarrow\infty$
for a nonlinear system, and the best estimate of $\hat \varphi(t)$ should decay
to unconditional mean of $\varphi$ (which is usually zero). 
The existence of a reduced Hamiltonian shows that this decay can happen only to a limited extent and thus 
the approximation can in general be valid only for short times. 
Equations (9) constitute the short time, 
or \lq\lq{first-order}", OP approximation.

Suppose however that instead of picking specific values for the initial data
$\hat x$
one samples them from  the invariant density $Z^{-1}e^{-\hat H(\hat x)}$.
The distribution of the $\hat x$'s is then invariant, and equal to their
marginal distribution in the full system (8) when the data are sampled from
the invariant distribution $Z^{-1}e^{-H(x)}$,
as one can  also see from  the identities
$\exp(-\hat H)=\exp \left(\log\int e^{-H}d\tilde x\right)=\int e^{-H}d\tilde x$.
The system (9) then generates the marginal probability density of part of the variables of
a system at equilibrium.
Thus OP at equilibrium is a way of 
reducing the number of variables 
without affecting the statistics of the variables that remain.
One can make short-time predictions about the future from the reduced system with coefficients
computed at equilibrium because it is self-consistent to assume for short
times that unresolved degrees of freedom
are in thermal equilibrium, as is explained in the OP papers cited above. 

\section{Renormalization}

For simplicity, we work here with real-space renormalization applied to
variables associated with specific sites in a plane, 
$x^{(1)}=(x_{I_1}, x_{I_ 2}, \ldots)$, where $I_k=(i_k,j_k)$,
$i_k, j_k$ are integers, all the $I_k$ are inside a square $D$ of side $N$ with $N$ large, and the
variables are listed in some convenient order.
The Hamiltonian $H=H^{(1)}$ is a function of $x^{(1)}$, $H^{(1)}=H^{(1)}(x^{(1)})$.
The need for the superscript (1) will appear shortly. 
We assume that the partition function $Z=\int e^{-H^{(1)}(x^{(1)})}dx^{(1)}$
is well defined, where $dx^{(1)}=dx^{(1)}_{I_1}dx_{I_2}^{(1)}\ldots$.

Suppose we group the variables $x_{I_1},x_{I_2},\ldots$ into groups
of $\ell$ variables (for example, we could divide $D$ into squares
each containing 4 variables).
The variables can be referred to as \lq\lq{spins}\rq\rq  in conformity with
common usage in physics.
Associate with each group a new variable $x^{(2)}_{J_1},x^{(2)}_{J_2},\ldots,$
where $J_1,J_2,\ldots$ is some ordering of the new variables and 
$x^{(2)}_{J_k}$ is a function (not necessarily invertible) of the 
$x^{(2)}_I$ in the group labeled by $J_k$, for example
$x^{(2)}_{J_k}=g(x^{(1)}_{I_{m+1}},
x^{(1)}_{I_{m+2}},\ldots, x^{(1)}_{I_{m+\ell}})$ for the appropriate $m$.
The vector $x^{(2)}$  is  
$x^{(2)}=(x^{(2)}_{J_1},x^{(2)}_{J_2}\ldots)$.
We can write
$$\matrix{Z&=&\int e^{-H^{(1)}(x^{(1)})} dx^{(1)}\hfill\cr
&&\cr
&=&\int dx^{(2)}
\int\delta \left(x^{(2)}-g(x^{(1)})\right) e^{-H^{(1)}(x^{(1)})}dx^{(1)}}.$$
where $dx^{(2)}=dx^{(2)}_{J_1}dx^{(2)}_{J_2}\cdots,$ and the $\delta$ 
function is a product of delta functions, one per group.
If one defines $H^{(2)}(x^{(2)})$ by the equation
$$e^{-H^{(2)}(x^{(2)})}=\int\delta\left(x^{(2)}-g(x^{(1)})\right)e^{-H(x^{(1)})}dx^{(1)},\eqno(10)$$
then $Z=\int e^{-H^{(2)}(x^{(2)})}dx^{(2)}$.

The mapping $x^{(1)}\rightarrow x^{(2)}$, followed by a change of numbering 
of the remaining variables so that $J_1, J_2\ldots$ (the indices of the new
variables $x^{(2)}$) enumerate  the new variables by going through all
integer pairs in a reduced domain of side $N/\sqrt\ell$, is a real-space
renormalization group transformation;
it produces a new set of variables which has less spatial detail than the 
previous set and such that distances between the remaining spins have been scaled down by $\sqrt\ell$.
If the calculation is set up so that the mapping $x^{(1)}\rightarrow x^{(2)}$,
$H^{(1)}\rightarrow H^{(2)}$ can be 
repeated, for example, if the range of the variables $x^{(1)}$ is invariant 
and the Hamiltonians $H^{(1)},H^{(2)}$ can be represented in the same finite-dimensional basis,
then one can produce in this way a sequence of Hamiltonians
$H^{(1)},H^{(2)},H^{(3)},\ldots$; 
the fixed points of the transformation 
$H^{(n)}\rightarrow H^{(n+1)}$ for a spin system of infinite spatial extent include the critical points of the system, 
see any discussion  of the RNG, for example \cite{kadanoff},\cite{ma}.

Consider the special case where $x_J^{(2)}$ is one of the  $x_I^{(1)}$
in its group--i.e., replace a block of spins by one of the spins in the block.
More general and widely used assignments of block variables will not be needed
in the present paper and will be discussed elsewhere.
We can identify the spins that remain with $\hat x$ of the
preceding section and the spins 
that disappear with $\tilde x$.
Equation (10) becomes a special case of equation (5),
and can be solved for $H^{(2)}$ by taking conditional
expectations of the derivatives of $H^{(1)}$.

Note that the usual RNG representation of a renormalized
Hamiltonian by means of 
additional couplings \cite{kadanoff} is interpreted here as an expansion of a conditional expectation 
in a  convergent series.
The new interpretation may be useful both in understanding what is
happening on the  computer and in deriving error estimates.
The relation between the RNG and conditional expectations shows that the 
latter can be calculated recursively, as we show in the example below.

We have written the RNG transformation above in notation suitable for spins
with a continuous range.
The case of discrete (e.g., Ising) spins is automatically included,
even though it may seem odd to differentiate  functions with a discrete 
domain and range. 
Indeed, add to the Hamiltonian $H$ a term of the form
$${\displaystyle 1\over\displaystyle\varepsilon}
\sum_i\left(\prod_j\psi(x_i-x_{0j})\right)$$
where $\varepsilon$ is small, the sum is over all spins, the product is over a 
finite number of values $x_{0j}$, and $\psi\geq0$ has a minimum at 0 and is positive elsewhere.
For small $\varepsilon$ such a term will constrain the $x_i$ to take on the
values $x_{0j}$, but since at the origin 
the derivative of $\psi$ is zero
the calculation of the conditional expectations is unaffected
by this term and the limit $\varepsilon\rightarrow0$ can
be taken without actually doing anything on the computer. All one has to do
is make sure that in the Monte-Carlo sampling only the values $x_{0j}$ are 
sampled. 
Indeed, results below will be given for Ising spins which take on the values $+1$ and $-1$,
with a \lq\lq{bare}'' (unrenormalized) Hamiltonian $H^{(1)}=\beta\sum x_Ix_J$, with summation over
locations $I,J$ that are neighbors on the lattice; $\beta=1/T$, where $T$ is the temperature.

\section{A decimation RNG/OP scheme for a spin system}

We consider in detail a RNG/recursive OP scheme where the number of
variables is halved at each step.
The spins are located on a square lattice with nodes $I_k=(i_k,j_k)$,
$i_k,j_k$ integers, and at each step of the recursion those for which 
$i_k+j_k$ is odd are eliminated while those for which $i_k+j_k$ is even
are kept. 
The spins with $i_k+j_k$ even constitute $\hat x$ and the others $\tilde x$;
the choice of which are even and which are odd is a matter of convention 
only (see Figure 1).
The variables are labeled by
$I_k: x_{I_1},x_{I_2},\ldots$.
\begin{figure}[ht]
\centerline{\scalebox{0.5}{\includegraphics{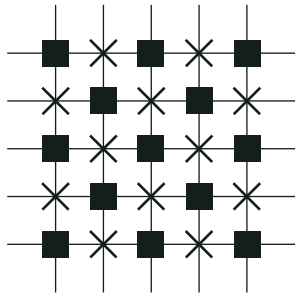}}}
\caption{The decimation pattern}
\label{chorin1.eps}
\end{figure}

Given a location $I=(i,j)$, we group the other variables according to their
distance from $I$:  group 1 contains only $x_I$, the variable at $I$. Group 2
(relative to $I$)
contains  those variables whose distance from $I$ is 1, group 3 contains
those variables whose distance to $I$ is $\sqrt 2$, etc.
We form the {\lq\lq}collective{\rq\rq} variables
$$X_{k,I}={\displaystyle 1\over\displaystyle n_k}\sum_{{\rm group\ }k} x_J$$
where $n_k$ is the number of variables in the group
(1 for group 1, 4 for group 2, etc.).
From these variables one can form a variety of translation-invariant 
polynomials in $x$ of various degrees:
$\sum_I x_I X_{k,I}=\sum_I X_{1,I}X_{k,I},$
$\sum_{I}(X_{k,I})^2(X_{k+1,I})^2,$
$\sum_I(X_{k,I})^4,\ldots$.
In practice the domain over which one sums must be finite, and it is natural 
to impose periodic boundary conditions at its edges to preserve the 
translation invariance.
We wrote out explicitly only polynomials of even degrees because the Hamiltonians
we consider are invariant under the transformation $x\rightarrow -x$.
The translation-invariant polynomials built up from the $X_{k,I}$ can be labeled
$\varphi_1 (x),\varphi_2(x),\ldots$
in some order.

Expand the $n$-th renormalized Hamiltonian in a series and keep the first
$\ell$ terms:
$$H^{(n)}=\sum_{k=1}^\ell \alpha_k^{(n)}\varphi_k(x).\eqno(11)$$
The derivative of this series at the spin $x_I$ is
$${\displaystyle \partial\over\displaystyle\partial x_I}
H^{(n)}=\sum_1^\ell \alpha_k^{(n)}\varphi_k^\prime(x),
\ \ \ \ \ \varphi_k^\prime=
{\displaystyle \partial\over\displaystyle\partial x_I}\varphi_k.
\eqno(12)$$
The functions $\varphi_k^\prime$ are easily evaluated, for example:
$$\left(\sum_J x_J X_{k,J}\right)^\prime=2 X_{k,I}$$
$$\left(\sum_J(X_{k,J})^4\right)^\prime=4\left(\sum_{{\rm group\ }k}x_J^3\right)/n_k^2,$$
etc., where \lq\lq{group} $k$" refers to distances from $I$, the variable with
respect to which we are differentiating (see Figure 2).

Pick a variable $x_I$ in $\hat x$
(i.e., $I=(i,j),$ $i+j$ even in our conventions).
Some of the functions  $\varphi_k^\prime$ in (12)  will be functions of $\hat x$
only and some will be functions of both $\hat x$ and $\tilde x$ or of 
$\tilde x$ only.
The task at hand is to project the latter on the former and then 
rearrange the series so as to shrink the scale of the physical domain.
To explain the construction 
we consider a very special case.

Suppose one can write
$$H^{(n)}(x)=\alpha_2^{(n)}\varphi_2+\alpha_3^{(n)}\varphi_3
+\alpha_4^{(n)}\varphi_4,\eqno(13)$$
where $\varphi_k(x)=\sum x_I X_{k,I}$ for $k=2,3,4,\ldots$.
Note that $\varphi^\prime_k={\partial\over\partial x_I}\varphi_k$
is a function only of $\hat x$ when $k=3,4$
(and when $k=6$, as we shall need to know shortly) but not when $k=2$ or 5
(see Figures 1, 2).
We now calculate the conditional expectations of the derivatives of $H^{(n)}$ given $\hat x$
by projecting them on the space of functions of $\hat x$.
First we project
$\varphi^\prime_2$
on the span of $\varphi^\prime_3,\varphi^\prime_4,\varphi^\prime_6$
(note that $\varphi^\prime_6$ is not in the original expansion (13)).
Form the matrix $\Phi$ with rows
$\langle \varphi_k^\prime,\varphi_3^\prime\rangle,
\langle \varphi_k^\prime,\varphi_4^\prime\rangle,
\langle \varphi_k^\prime,\varphi_6^\prime\rangle$
for $k=3,4,6$, the primes once again denoting differentiation with respect
to $x_I$.
Form the vector $r$ with component
$\left(\langle \varphi_2^\prime,\varphi_3^\prime\rangle,
\langle \varphi_2^\prime,\varphi_4^\prime\rangle,
\langle \varphi_2^\prime,\varphi_6^\prime\rangle\right)$.
Let $c=(c_1,c_2,c_3)$ be the solution of $\Phi c=r$ (see equation (4)).
The coefficients $c$ are the coefficients of the orthogonal projection of 
$\varphi^\prime_2$ onto the span of  $\varphi^\prime_3,\varphi^\prime_4,\varphi^\prime_6$
which is contained in $\hat L_2$.
After projection, 
the coefficients of  $\varphi_3,\varphi_4$ in (13) become
$$\alpha_3^{{\rm new}}=\alpha_3^{(n)}+\alpha_2^{(n)}c_1,$$
$$\alpha_4^{{\rm new}}=\alpha_4^{(n)}+\alpha_2^{(n)}c_2,$$
and $\varphi_6$ acquires the coefficient $\alpha_2^{(n)}c_3$.

\begin{figure}[ht]
\centerline{\scalebox{0.5}{\includegraphics{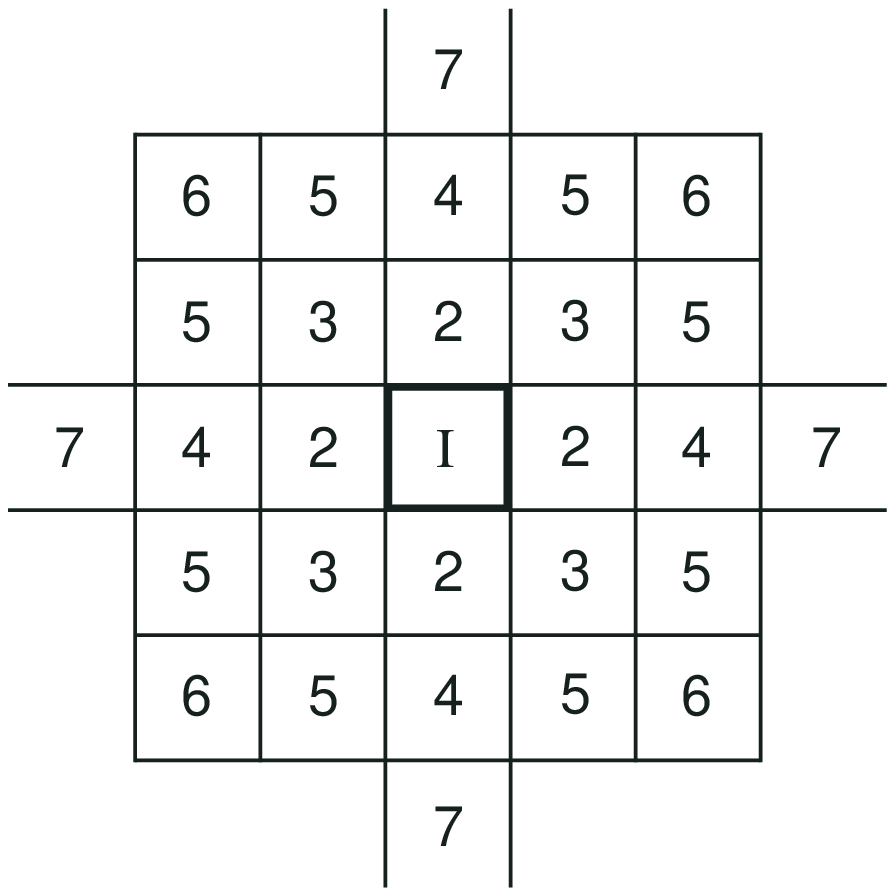}}}
\caption{The collective variables}
\label{chorin2.eps}
\end{figure}
If one relabels the remaining spins so that they occupy the lattice
previously occupied by all the spins, group 3 becomes group 2, group 4
becomes group 3, and group 6 becomes group 4 (see Figure 2).
The new Hamiltonian $H^{(n+1)}$ now has the representation 
$$H^{(n+1)}=\alpha_2^{(n+1)}\varphi_2+\alpha_3^{(n+1)}\varphi_3
+\alpha_4^{(n+1)}\varphi_4,$$
with
$$\matrix{
\alpha_2^{(n+1)}&=&\alpha_3^{(n)}+\alpha_2^{(n)}c_1,\hfill\cr
&&\cr
\alpha_3^{(n+1)}&=&\alpha_4^{(n)}+\alpha_2^{(n)}c_2,\hfill\cr
&&\cr
\alpha_4^{(n+1)}&=&\alpha_2^{(n)}c_3.
}$$
More generally, if $H^{(n)}$ is expressed as a truncated series, partition the terms
in the series for ${\partial\over\partial x_I}H^{(n)}$ into functions
of $\hat x$ and functions of both $\hat x$ and $\tilde x$.
Add to the terms which are functions of $\hat x$ additional terms which
are functions of $\hat x$ and are chosen so that after relabelling they acquire the form of terms
already in the series (just as above, terms that depend on $X_{3,J}$,
for example, become terms that depend on $X_{2,J}$ after 
relabelling). 
Project the functions of $x$ on the span of the expanded set of functions of $\hat x$, collect terms and relabel.
This is a renormalization step, and it can be repeated.

Note that it if one wants to reduce the number of variables by a given factor, one can in principle
use an analogous RNG/conditional expectation construction and get there in one iteration; the
recursive construction is easier to do and the intermediate Hamiltonians, whose coefficients constitute
the parameter flow in the renormalization, contain useful information. 

The discussion so far may suggest that one sample the Hamiltonians recursively,
i.e., start with $H^{(1)}$,
find $H^{(2)}$, use Monte-Carlo to sample the density $Z^{-1}e^{-H^{(2)}}$ and
find $H^{(3)}$ etc. 
The disadvantages of this approach are: (i) The sampling of the densities
$Z^{-1}e^{-H^{(n)}}$
can be much more expensive for $n>1$ than for $n=1$ because each proposed Monte-Carlo move
may require that the full series for
${\partial\over\partial x_I}H^{(n)}$
be summed twice; and (ii) each evaluation of a new
Hamiltonian is only approximate because the series are truncated, and, more important, the Monte-Carlo
evaluation of the coefficients may have limited accuracy. These errors accumulate from step to step
and may produce false fixed points and other artifacts. 

The remedy lies in Swendsen's observation \cite{burkhardt},\cite{swendsen} that the successive Hamiltonians can be 
sampled without being known explicitly. Sample the original Hamiltonian, remove the unwanted spins
and
relabel the remaining spins so as to cover the original lattice, as in the relabelling step in the
renormalization; the probability density of the remaining spins is
$Z^{-1}e^{-H^{(2)}}$; 
repeating $n$ times yields samples of $Z^{-1}e^{-H^{(n+1)}}$.
The price one pays is that to get an
$m$ by $m$ sample of $Z^{-1}e^{-H^{(n)}}$
one has to start by sampling a $2^qm$ by $2^qm$ array 
of non-renormalized spins, where $q$ is either $(n+1)/2$ or $n/2$ depending on the parity of $n$ and on programming choices; 
the trade-off is in general very worthwhile.  What has been
added to Swendsen's calculation is an effective evaluation of the coefficients of the expansion of
$H^{(n)}$ from the samples. 

The programming here requires some care. With the decimation scheme as in Figure 1, after one removes
the unwanted spins in $x^{(n)}$ the remaining spins, the variables $x^{(n+1)}$, live on a lattice with a mesh size $\sqrt2$ larger than before;
after relabelling they find themselves on a lattice with the same mesh size as before but arranged at a $\pi/4$ angle
with respect to 
the previous lattice. To extract a square array from at this set of spins one has to make the size of the box
that includes all the spins half the size of the previous box. At the next renormalization one obtains $x^{(n+2)}$ 
which can be extracted from $x^{(n)}$ by taking one spin in four and the resulting box size is the same
as the size of the box that contains $x^{(n+1)}$. One may worry a little about boundary conditions for $x^{(n+1)}$:
the periodicity of $x^{(n)}$ is not the same as the periodicity one has to assume for $x^{(n+1)}$ because of the
rotation; the resulting error is too small to be detected in our calculations.

\section{Some numerical results}

We now present some numerical results obtained
with the RNG/conditional expectation scheme.  
The problem we apply the construction to is Ising spins; more interesting 
applications will be presented elsewhere. The point being made is that the construction
can be effectively implemented. 
The results are presented for Ising spins.

In table I we list the coefficients $\alpha_k^{(n)}$ in the expansion of
$H^{(n)} $ for $n=1,\ldots7$ and $T=2.27$.
The functions $\varphi_k$ are as follows:
$$\varphi_k=\sum x_J X_{k,J} \ \ \ {\rm for}\ k=1,2,3,4,5,6$$
$$\varphi_{6+k}=\sum (X_{k+1,J})^4,\ \ \ {\rm for}\ k=1,2,3$$
$$\varphi_{10}=\sum X_{2,J}^2 X_{3,J}^2.$$
Note that as a result of the numbering of the $\varphi$'s the last coefficient is not
necessarily the smallest coefficient. 
This table represents the parameter flow and if the functions $\varphi_k$ are written in terms of the
variables $x_J$ the table defines the new system of equations for the reduced set of variables.
Remember that in the projection on $\hat L_2$ additional functions are used
so that after relabelling the series has the same terms , but maybe with different coefficients, as
before the renormalization. In $H^{(1)}$, $\alpha_2$ is the sole non-zero coefficient, 
and its value is determined by $T$ and the definition of $X_{2,J}$, in particular the
presence of the coefficient $n_2$ (see above).  

It is instructive to use the parameter flow to identify the critical 
temperature $T_c$.
For $T<T_c$ the renormalization couples ever more distant spins while
for $T>T_c$
the spins become increasingly decoupled.
One can measure the increasing or decreasing coupling by considering the
quadratic terms in the Hamiltonian (the terms of the form $\sum x_J X_{k,J}$)
and calculating the \lq\lq{second moments}" $M_2$ of their
coefficients $\alpha_k^{(n)}$:
$$M_2^{(n)}=\sum_{k=2}^\ell d_k^2 \alpha_k^{(n)}$$
where $d_k$  is the distance from $J$ of the spins in the group $k$
(see the definition of $X_{k,J}$),  
$\alpha_k^{(n)}$ is the coefficient of $\sum x_J X_{k,J}$ in the expansion
of $H^{(n)}$, and $\ell$ is the number of  quadratic terms in this expansion.
In Figure 3 we show the evolution of $M_2^{(n)}$ with $n$  for various values of $T$ 
(with $\ell=5$ and 7 functions over-all in the expansion, including
non-quadratic functions).
\begin{figure}[ht]
\centerline{\scalebox{0.5}{\includegraphics{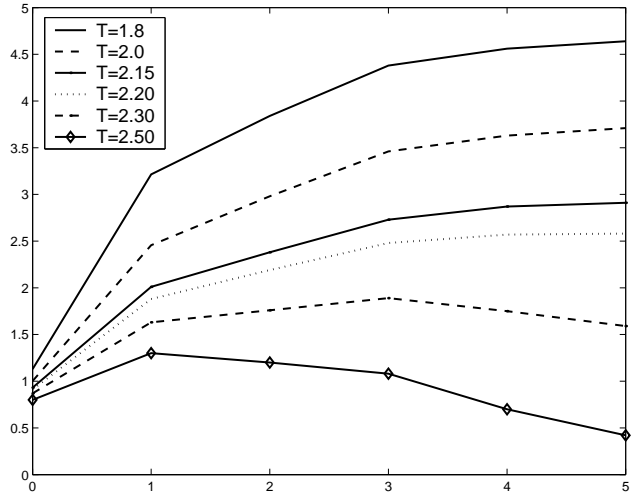}}}
\caption{Second moments of the coefficients of the renormalized
Hamiltonian for various values of $T$ for successive iterations}
\label{chorin3.eps}
\end{figure}

In Figure 4 we show the evolution of $M_2^{(n)}$ near
$T_c=2.269\ldots$ with $\ell=6$ and  10 terms in the expansion.
The non-uniform behavior of $M_2$ is not a surprise
(it is related to the non-uniform convergence
of critical exponents  already observed by Swendsen).
Each step in the renormalization used $10^5$ Monte-Carlo steps per spin.
From  these graphs one would conclude that $T_c\sim2.26$, an error of .5\%.
The accuracy depends on the number of terms in the expansion and on the choice of terms; with only 6 terms (4 quadratic
and 2 quartic), the error in the location of $T_c$ increases to about 3\%. 
The point is not that this is a good way to find $T_c$ but that it is a check
on  the accuracy of the parameter flow. From the Table one can see that the system first approaches
the neighborhood of a fixed point and then diverges from it, as one should expect in a discrete 
sequence of transformations. 

\begin{figure}[ht]
\centerline{\scalebox{0.5}{\includegraphics{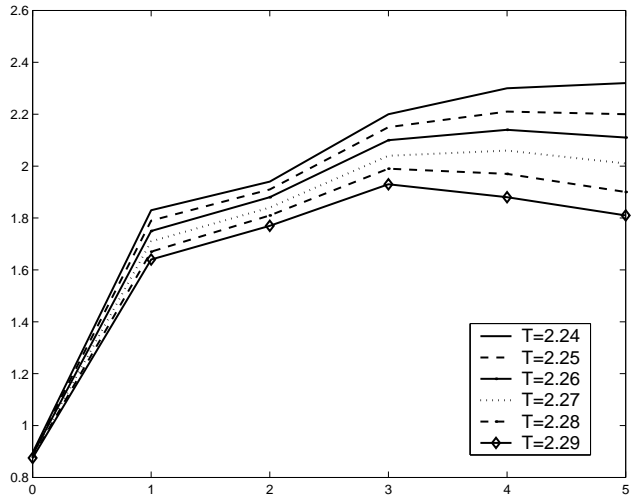}}}
\caption{Second moments of the coefficients of the renormalized Hamiltonian near $T_c$}
\label{chorin4.eps}
\end{figure}

\vskip14pt
$$\begin{tabular}{ccrrrrrr}
\multicolumn{8}{c}{Table 1}\\
\multicolumn{8}{c}{Parameter flow for the Ising model $T=2.26$, 10 basis functions}\\
\hline
{\rm iteration}&1&2&3&4&5&6&7\\[0.5ex]
$\alpha_1$&0&.26&.35&.44&.48&.52&.54\\
$\alpha_2$&.893&.47&.47&.35&.30&.25&.21\\
$\alpha_3$&0&.32&.20&.23&.21&.20&.18\\
$\alpha_4$&0&.04&.08&.11&.12&.13&.13\\
$\alpha_5$&0&.07&.11&.13&.13&.12&.12\\
$\alpha_6$&0&$-$.01&.01&.01&.02&.03&.02\\
$\alpha_7$&0&$-$.08&$-$.07&$-$.10&$-$.09&$-$.09&$-$.08\\
$\alpha_8$&0&.04&.02&.02&.01&.00&$-$.10\\
$\alpha_9$&0&$-$.00&$-$.01&$-$.00&$-$.00&.00&.00\\
$\alpha_{10}$&0&$-$.12&$-$.17&$-$.18&$-$.18&$-$.17&$-$.16
\end{tabular}
$$
\begin{figure}[ht]
\centerline{\scalebox{0.5}{\includegraphics{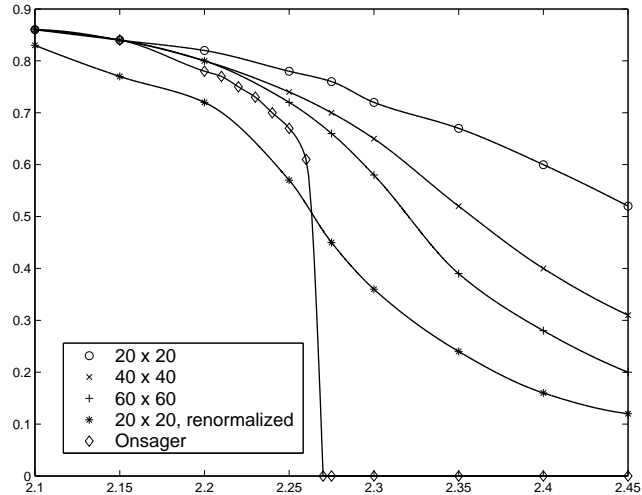}}}
\caption{Bare and renormalized magnetization near $T_c$}
\label{chorin5.eps}
\end{figure}

We now use the renormalized system to calculate the magnetization
$m=E[\sum x_I/n^2]$.  
To get the correct non-zero
$m$ for $T<T_c$ on a small lattice the symmetry must be broken, and we do this by imposing on all the arrays
the boundary condition $x_{boundary}=1$ rather than the periodic boundary conditions
used elsewhere in this paper. In Figure 5 we display $m$ computer with the bare (unrenormalized) Hamiltonian
$H^{(1)}$ on 3 lattices: $20$ by $20$, $40$ by $40$, $60$ by $60$, as well as the results obtained on a $20$ by
$20$ lattice  
by sampling the density defined by the renormalized Hamiltonian $H^{(5)}$ which corresponds in principle
to an $80$ by $80$ bare calculation. We also display the exact Onsager values of $m$.  The calculations focus on values of $T$ in the neighborhood of $T_c$ where the
size of the lattice matters; one cannot expect the  results to agree perfectly with the Onsager results on a finite lattice with periodic boundary conditions for any $n$; all one can expect is to have the values of the small renormalized calculation be consistent with 
results of a larger bare calculation.  
We observe that they do, up to the shift in $T_c$ already pointed out and due to the choice of basis functions. 

The determination of the critical exponents 
for a spin model is independent of the determination of the coefficients in the 
expansion of $H^{(n)}$, and is mentioned here only because it does provide 
a sanity check on the constructions, in particular on the adequacy of the basis functions. 
For comparable earlier calculations, see in particular Swendsen's chapter in \cite{burkhardt}.
As is well known,
if $A$ is the matrix of derivatives
$\partial \alpha_i^{(n+1)}/ \partial\alpha_j^{(n)}$
at $T=T_c$, those of its eigenvalues that are larger than 1 are the critical
exponents of the spin system \cite{kadanoff}.
The matrix $A$ can be found from the chain rule \cite{binney},\cite{burkhardt}
$${\displaystyle\partial\over\displaystyle \partial\alpha_j^{(n)}}
E\left[\varphi_k\left(x^{(n+1)}\right)\right]=\sum_i
{\displaystyle\partial\alpha_i^{(n+1)}\over\displaystyle \partial\alpha_j^{(n)}}
\  {\displaystyle\partial E[\varphi_k(x^{(n+1)})]\over\displaystyle \partial\alpha_i^{(n+1)}}$$
and the sum is over all the coefficients that enter the expansion.
The derivatives of the expectations are given by correlations as follows:
$${\displaystyle\partial E[\varphi_k(x^{(n+1)})]\over\displaystyle \partial\alpha_j ^{(n)}}
=E\left[\varphi_k(x^{(n+1)})\varphi_j(x^{(n)})\right]-
E[\varphi_k(x^{(n+1)})]E[\varphi_j(x^{(n)})],$$
$${\displaystyle\partial E[\varphi_k(x^{(n+1)})]\over\displaystyle \partial\alpha_i ^{(n+1)}}
=E\left[\varphi_k(x^{(n+1)})\varphi_i(x^{(n+1)})\right]-
E[\varphi_k(x^{(n+1)})]E[\varphi_i(x^{(n+1)})],$$
see \cite{burkhardt}. 
In most of the literature on real-space renormalization for Ising spins the 
variables $x^{(n+1)}$ are obtained from  $x^{(n)}$ by \lq\lq{majority rule}", i.e., 
by assigning to the group that defines $x^{(n+1)}$ the value $+1$
if most of the members of the group are $+1$, the value $-1$ if most of the
members of the group are $-1$, with ties resolved at random.
For the decimation scheme described above our 
\lq\lq{pick one}" rule ($x^{(n+1)}$ is one of the members of the group) is identical 
to the majority rule.
There is an apparent difficulty  in the decimation because at each recursion the number
of terms in the summation that defines 
the basis functions is reduced by half while the square root of an integer is
not in general an integer, so that one has to perform
Swendsen sampling on rectangles so designed that the ratio of the areas of two successive rectangles is 1/2.  This has not turned out
to be harmful,
and the value of $\nu$, the correlation exponent, was found to be 
$1$ (the exact value) $\pm.01$ with $10^6$ Monte-Carlo moves per spin, the error depending mainly on the number of Monte-Carlo
moves which has to be very large, in line with previous experience \cite{swendsen}. 
We also checked that in a renormalization scheme where a $2\times2 $ block
of spins is replaced at each iteration by a single spin, the \lq\lq{majority rule}"
and our \lq\lq{pick one}" rule  for $x^{(n+1)}$ yield similar results.
One needs fewer terms in the expansion of the Hamiltonian to get accurate
values of the exponents than to get an accurate parameter flow, but a larger number of Monte-Carlo moves. 

\section{Conclusions}
We have presented a simple relation between conditional
expectations for systems at equilibrium
on one hand and the RNG on the other,
which makes it possible to find
efficiently the coefficients in a reduced systems of equations for a subset of variables whose
distribution as given by reduced system equals their marginal distribution in the original system. 
The  numerical results above emphasized the neighborhood of the critical point in the simple
example because this is where the variables are strongly coupled without separation of scales 
and a reduction in system size requires non-trivial tools.
The next steps will be the application of these ideas to time-dependent
problems and to finite-difference approximations of underresolved partial differential 
equations, along the lines suggested in \cite{goldenfeld}; this work will be presented elsewhere.

\vskip14pt
\noindent
{\bf Acknowledgments.}
I would like to thank Prof. G.I.\ Barenblatt, 
Prof. N.\ Goldenfeld,
Prof. O.\ Hald,
Prof. R.\ Kupferman,
Mr.\ K.\ Lin, and
Mr.\ P.\ Stinis
for very helpful discussions and comments.
This work was supported in part by the Office of Science,
Office of Advanced Scientific Computing Research,
Mathematical, Information, and Computational Sciences Division,
Applied Mathematical Sciences Subprogram,
of the U.S.\ Department of Energy under Contract No.\ DE-AC03-76SF00098
and in part by the National Science Foundation under grant number DMS89-19074.
\vfill\eject


\end{document}